\documentclass[12pt,preprint]{aastex}
\usepackage{amsmath,amssymb}
\usepackage{graphicx}
\usepackage{hyperref}

\slugcomment{KSUPT-11/5 \hspace{0.5truecm} October 2011}
\def\intl{\int\limits}

\begin{document}
\title{Constraints on dark energy from H II starburst galaxy apparent magnitude versus redshift data}
\author{Data Mania\altaffilmark{1,2} and Bharat Ratra\altaffilmark{1}}

\altaffiltext{1}{Department of Physics, Kansas State University, 
                 116 Cardwell Hall, Manhattan, KS 66506, USA \ 
                 mania, ratra@phys.ksu.edu}
\altaffiltext{2}{Center for Elementary Particle Physics, 
                 Ilia State University, 3-5 Cholokashvili Ave., 
                 Tbilisi 0179, Georgia}

\begin{abstract}
In this paper we use H II starburst galaxy apparent magnitude versus 
redshift data from \cite{siegel2005} to constrain dark energy cosmological 
model parameters. These constraints are generally consistent with those 
derived using other data sets, but are not as restrictive as the tightest 
currently available constraints.
\end{abstract}
%\maketitle

\section{Introduction}

There is significant observational evidence supporting the idea that we
live in a spatially-flat Universe that is currently undergoing accelerated 
cosmological expansion. Most cosmologists believe that this accelerated 
expansion is powered by dark energy which dominates the current cosmological
energy budget \citep[for reviews of dark energy see][and references therein]
{tsujikawa10, linder10, blanchard10, sapone10}.\footnote[3]{
Some, however, feel that this accelerated expansion is better viewed 
as an indication that the general relativistic description of gravitation
needs to be improved on \citep[see][and references therein]{defelice10, 
tsujikawa10, jain10}. In this paper we assume 
that general relativity provides an adequate description of gravitation
on cosmological scales.}

The energy budget of the ``standard'' model of cosmology --- the 
spatially-flat $\Lambda$CDM model \citep{peebles84} --- is currently 
dominated by far by
dark energy: Einstein's cosmological constant $\Lambda$ contributes more 
than 70 \% of the budget. Nonrelativistic cold dark matter (CDM) is the next 
largest contributor (more than 20 \%), followed by nonrelativistic baryons
(around 5 \%). For a review of the standard model see \cite{ratra08} and 
references therein. It has been know for some time that the 
$\Lambda$CDM model is reasonably consistent with most observational 
constraints \citep[see, e.g.,][for early indications]{jassal10, wilson06, Davis2007, allen08}.\footnote[4]{
The $\Lambda$CDM model assumes the ``standard'' CDM structure formation
picture, which might be in some observational difficulty \citep[see, e.g.,][] 
{Peebles&Ratra2003, Perivolaropoulos2010}.}   

However, the $\Lambda$CDM model appears to leave some conceptual questions 
unanswered. One of these is that the naive energy density scale we expect 
from quantum field theory considerations is many orders of magnitude higher 
than the measured cosmological constant energy density scale. Another is 
what is known as the coincidence puzzle: The energy density of a cosmological 
constant does not change over time but the matter density decreases with the 
cosmological expansion, so it is puzzle why we observers live at this 
seemingly special time, when the dark energy and nonrelativistic matter 
densities happen to be of comparable size.

These puzzles, and perhaps others, could be alleviated if we assume that 
the dark energy density was higher in the past and slowly decreased in time, 
thus remaining comparable to the nonrelativistic matter density for a longer 
time \citep{Ratra&Peebles1988}. Many such time-varying dark energy models 
have been proposed.\footnote[5]{
For recent discussions see, e.g., \cite{Novosyadlyj2010}, \cite{harko10}, 
\cite{keresztes10}, \cite{pettorino10}, \cite{liu10}, \cite{costa10}, 
\cite{farajollahi10}, and references therein.}
In this paper, for illustrative purposes, we consider two dark energy models 
and one dark energy parametrization. 

In the $\Lambda$CDM model, time-independent dark energy ---
the cosmological constant --- may be modeled as a spatially
homogeneous fluid with equation of state parameter
$\omega_\Lambda = p_\Lambda/\rho_\Lambda = -1$ (where $p_\Lambda$
and $\rho_\Lambda$ are the fluid pressure and energy density).

In an attempt to describe slowly decreasing dark energy, 
the dark energy fluid equation of state is extended to the XCDM 
$X$-fluid parametrization. Here, dark energy is again modeled 
as a spatially homogeneous ($X$) fluid, but now with an equation 
of state parameter $w_X = p_X/\rho_X$, where $w_X (< -1/3)$ is 
an arbitrary constant and $p_X$ and $\rho_X$ are the pressure 
and energy density of the $X$-fluid. When $w_X = -1$ the XCDM 
parametrization reduces to the complete and consistent 
$\Lambda$CDM model. For any other value of $w_X (< -1/3)$, the 
XCDM parametrization is incomplete as it cannot describe spatial 
inhomogeneities \citep[see, e.g.][]{ratra91, podariu2000}. For 
computational simplicity, here we study the XCDM parametrization 
in only the spatially-flat cosmological case.

A slowly rolling scalar field $\phi$, with decreasing (in $\phi$)
potential energy density $V(\phi)$, can be used to consistently 
model a gradually decreasing, time-evolving, dark energy density.
An inverse power-law potential energy density $V(\phi) \propto 
\phi^{-\alpha}$, where $\alpha$ is a non-negative constant, 
provides a simple realization of this $\phi$CDM scenario  
\citep[][]{Peebles&Ratra1988, Ratra&Peebles1988}.  When $\alpha = 0$ 
the $\phi$CDM model reduces to the corresponding $\Lambda$CDM case. 
For computational simplicity, here we study the $\phi$CDM model
in only the spatially-flat cosmological case.

Current observational data convincingly indicate that the 
cosmological expansion is accelerating. The main support for 
accelerated expansion comes from three types of data: supernova 
Type Ia (SNIa) apparent magnitude versus redshift measurements
\citep[see, e.g.,][]{Amanullah2010, holsclaw10, wangetal11, guy10}; 
cosmic microwave background (CMB) anisotropy data 
\citep[see, e.g.,][]{Podariu2001b, Komatsu2009, liu11, lavacca11} 
combined with low estimates of the cosmological matter density 
\citep[see, e.g,][]{chen03b}; and, baryon acoustic oscillation (BAO) 
peak length scale estimates \citep[see, e.g.,][] {samushia09, wang09, 
Percival2010, blake11}. However,
the error bars associated with these three kinds of data are still 
too large to allow for a significant observational discrimination 
between the $\Lambda$CDM model and the two simple time-varying dark
energy models discussed above.  
 
There are two major reasons to consider additional data sets. 
First, it is of great interest to check the above results by comparing
to results derived from other data. If the new constraints differ 
significantly from the old ones, this could mean that the model being
tested is observationally inconsistent, or it could mean that one
of the data sets had an undetected systematic error. Either of these
is a significant result. On the other hand, if the constraints from the
new and the old data are consistent, then a joint analysis 
of all the data could result in significantly tighter constraints, and
so possibly result in significantly discriminating between constant and 
time-varying dark energy.

Other data that have recently been used to constrain dark energy models
include strong gravitational lensing measurements \citep[e.g.,][]{chae04, 
lee07, Zhang2010, biesiada10}, angular size as a function of redshift 
observations \citep[e.g.,][]{guerra00, chen03a, Bonamente2006}, 
Hubble parameter as a function of redshift measurements 
\citep[e.g.,][]{Jimenezetal2003, Samushia&Ratra2006, Sen&Scherrer2008, 
samushia07, Panetal2010}, galaxy cluster gas mass fraction data
\citep[e.g.,][]{allen08, Samushia&Ratra2008, ettori09, tong11}, and 
large-scale structure observations \citep[e.g.,][]{baldiandpettorino10, 
deboni10, mortonson2011, brouzakis11, campanelli11}.

While constraints from these data are less restrictive than those 
derived from the SNeIa, CMB and BAO data, both types of data result 
in largely compatible parameter restrictions that generally support 
a currently accelerating cosmological expansion. This gives us confidence 
that the broad outlines of the ``standard'' cosmological model are now 
in place. However, there is still ambiguity. Current observations are
unable to differentiate between different dark energy models. For instance,
while current data favor a time-independent cosmological constant, they are
unable to rule out time-varying dark energy. More and higher-quality data
is needed to accomplish this important task.

It is anticipated that future space missions will result in significantly 
more and better SNeIa, BAO, and CMB anisotropy data 
\citep[see, e.g.,][]{podariu01a, samushia11, wang10, astier2010}.
A complementary approach is to develop cosmological tests that
make use of different sets of objects. Recent examples include
the lookback time test \citep[e.g.,][]{pires06, dantas11}
and the gamma-ray burst luminosity versus redshft test 
\citep[see, e.g.][]{schaefer07, wang08, XuandWang2010}.
Gamma-ray bursts, in particular, are very luminous and can be seen to 
much higher redshift than the SNeIa and so probe an earlier cosmological epoch.
H II starburst galaxies are also very luminous and can be seen to redshifts 
exceeding 3. Recent work has indicated that H II starbusrt galaxies might 
be standardizable candles \citep{melnick00, terlevich02, melnick2003star}, 
because of 
the correlation between their velocity dispersion, $H_\beta$ luminosity,
and metallicity \citep{melnick78, terlevich81, melnick88}.

In this paper we use H II galaxy data from \cite{siegel2005} to constrain 
parameters of the three dark energy models mentioned above. 
\citet{plionisetall09, plionisetal10, plionisetal11} have used the 
\citet{siegel2005} data to constrain the XCDM parametrization.
Here we also constrain parameters of the consistent and physically
motivated $\Lambda$CDM and $\phi$CDM cosmological models. We also
derive constraints on the parameters of these models and the 
XCDM parametrization from a joint analysis of the \citet{siegel2005}
H II galaxy data and the \citet{Percival2010} BAO peak length scale 
measurements.

In the next section we list the basic equations for the dark energy 
models we consider. In Sec.\ \ref{hiic} we discuss the analysis method 
used for the H II galaxy data and the resulting constraints and how they
compare with those from other cosmological tests. Section \ref{joint} 
presents constraints from a joint analysis of H II and BAO data. Section 
\ref{conc} summarizes our results.

\section{Dark energy model equations}
\label{dee}

The Friedman equation for the $\Lambda$CDM model with spatial 
curvature is
\begin{equation}
H(z, H_0, {\bf p})=H_0\sqrt{\Omega_m(1+z)^3+\Omega_\Lambda+\Omega_R(1+z)^2}.
\end{equation}
Here $H_0$ is the Hubble constant and $H(z)$ is the Hubble parameter at 
redshift $z$, $\Omega_m$, $\Omega_\Lambda$, and 
$\Omega_R=1-\Omega_m-\Omega_\Lambda$ are the nonrelativistic (baryonic 
and cold dark) matter, cosmological constant, and space curvature energy 
density parameters, respectively, and the model parameter set ${\bf p} = 
(\Omega_m, \Omega_\Lambda)$. 

For the spatially-flat XCDM parametrization, the Hubble parameter is
\begin{equation}
H(z, H_0, {\bf p})=H_0\sqrt{\Omega_m(1+z)^3+(1-\Omega_m)(1+z)^{3(1+w_X)}}.
\end{equation}
Here $w_X < -1/3$ and the model parameter set ${\bf p} = 
(\Omega_m, \omega_X)$. 

For the spatially-flat $\phi$CDM case we have a coupled set of equations 
governing the scalar field and Hubble parameter evolution
\begin{equation}
\ddot\phi+3H\dot\phi-\dfrac{\kappa\alpha}{2G}\phi^{-(\alpha+1)}=0, 
\end{equation}
%3
%4
\begin{equation}
H(z, H_0, {\bf p})=H_0\sqrt{\Omega_m(1+z)^3+\Omega_\phi(z)} .
\end{equation}
Here we consider the inverse power law scalar field $(\phi)$ potential
energy density \citep{Peebles&Ratra1988} so 
\begin{equation}
\Omega_\phi(z)=\dfrac{1}{12H_0^2}\left(\dot\phi^2+\dfrac{\kappa}{G}
\phi^{-\alpha}\right),
\end{equation}
where $G$ is the Newtonian gravitational constant and $\alpha > 0$ is a 
free parameter (that determines $\kappa$). In this case the
model parameter set is ${\bf p} = (\Omega_m, \alpha)$.

The distance modulus as a function of redshift is
\begin{equation}
\mu(z, H_0, {\bf p})=25+5\log\left[3000(1+z)y(z)\right]-5\log h,
\end{equation}
where $h$ is the dimensionless Hubble constant defined from 
$H_0=100 h \mbox{ km s}^{-1}~ \mbox{Mpc}^{-1}$. In the spatially
open case $y(z)$ is 
\begin{equation}
y(z, {\bf p})=\frac{1}{\sqrt{\Omega_R}}
\sinh\left(\sqrt{\Omega_R}\intl_0^z \frac{dz'}{E(z')}\right), 
\end{equation}
where $E(z) = H(z)/H_0$. When $\Omega_R=0$ this reduces to 
$
y(z)=\intl_0^z {dz'/E(z')}.
$ 
The angular diameter distance $d_A = cy(z)/[H_0 (1 + z)]$.

\section{Constraints from H II galaxy apparent magnitude data}
\label{hiic}

We use the 13 $\mu_{\rm obs} (z_i)$ measurements of \citep{siegel2005},
listed in \hyperref[tabl]{Table 1}, to constrain cosmological parameters by 
minimizing
\begin{equation}
\chi^2_{\rm H II} (H_0, {\rm p}) = \sum^{13}_{i=1} {\left[\mu_{\rm obs}(z_i) - 
\mu_{\rm pred}(z_i,H_0,{\bf p})\right]^2 \over \sigma_i^2}.
\end{equation}
Here $z_i$ is the redshift at which $\mu_{\rm obs}(z_i)$ is measured, 
$\mu_{\rm pred}(z_i,H_0,{\bf p})$ is the predicted distance modulus at
the same redshift for the model under consideration, and $\sigma_i$ is
the average of the upper and lower error bars listed in \hyperref[tabl]{Table 1}. 

The \citet{siegel2005} measurements listed in \hyperref[tabl]{Table 1} are computed from 
\begin{equation}
\mu_{\rm obs}=2.5\log\left(\dfrac{\sigma^5}{F_{H\beta}}\right) - 
2.5\log\left(\dfrac{O}{H}\right)-A_{H\beta}+Z_0
\end{equation}
where $F_{H\beta}$ and $A_{H\beta}$ are the $H_\beta$ flux and extinction
and $O/H$ is a metallicity. 
Following \cite{plionisetal11}, for the zero point magnitude we use 
$Z_0=-26.60$, use Hubble constant value $H_0=73 \mbox{ km s}^{-1}~ 
\mbox{Mpc}^{-1}$ (and do not account for the associated uncertainty),
and also exclude two H II galaxies (Q1700-MD103 and SSA22a-MD41) 
that show signs of a considerable rotational velocity component 
\citep{erb2006}. 

The best fit parameter set $\bf p_*$ is where $\chi^2_{\rm H II}$ is 
at a minimum, $\chi^2_{\rm min}$. The $1\sigma$, $2\sigma$, and 
$3\sigma$ confidence intervals, in the two-dimensional parameter 
space, are enclosed by the 
contours where $\chi^2=\chi^2_{\rm min} + \Delta\chi^2$ with  
$\Delta\chi^2=2.30$, $\Delta\chi^2=6.17$, and $\Delta\chi^2=11.8$, 
respectively.

The H II galaxy data constraints on cosmological parameters are shown 
in Figs.\ 1---3. Our results generally agree with \cite{plionisetal11} 
for the XCDM parametrization (compare our \hyperref[fig: 2]{Fig. 2} and their Fig.\ 10).
The small differences are due to the fact that in our analysis we do 
not use specially weighted sigmas, but rather average distance moduli 
uncertainties, and also ignore gravitational lensing effects.
Both analyses ignore small $H_0$ uncertainties; these are not important
for our illustrative purposes here, but should be accounted for in an
analysis of improved near-future H II galaxy data. 

The H II galaxy data constraints in Figs.\ 1---3 are not as restrictive
as those that follow from SNeIa, BAO, or CMB anistropy data. They
are, however, comparable to those from Hubble parameter observations 
\citep[see][and references therein]{Chen2011b} or lookback time observations
\citep[see][and references therein]{Samushiaetal2010}, and somewhat
more restrictive than angular diameter distance constraints 
\citep[see][and references therein]{Chen2011a} and gamma-ray burst
luminosity distance ones 
\citep[see][and references therein]{Samushia&Ratra2010}. We again
note that we have not accounted for the uncertainty in $H_0$ in our
analysis, so the H II galaxy constraints derived here are a little 
more restrictive than they really are. However, our analysis
clearly illustrates the potential constraining power of near-future 
H II galaxy luminosity distance data.

\section{Joint constraints from BAO and H II galaxy data}
\label{joint}

To constrain observables using BAO data we follow the procedure of 
\cite{Percival2010}. With $D_V (z) = [(1 + z)^2 d_A^2 cz/H(z)]^{1/3}$,
\cite{Percival2010} measure
\begin{equation}
D_V(0.275)=(1104\pm30)\left(\dfrac{\Omega_bh^2}{0.02273}\right)^{-0.134}\left(\dfrac{\Omega_mh^2}{0.1326}\right)^{-0.255} \mbox{~Mpc} .
\end{equation}
We use this to build the likelihood estimator $\mathcal{L}_{\rm BAO}$ with a 
Gaussian prior of $\Omega_m h^2=0.1326\pm0.0063$, and neglect the error 
for $\Omega_b h^2$ as WMAP5 data constrains it to 0.5 \% \citep{Komatsu2009}. 
The BAO data constraint contours are shown in Figs.\ 4---6.

To derive joint H II galaxy and BAO data constraints we maximize the 
products of likelihoods $\mathcal{L}({\bf p}) = \mathcal{L}_{\rm {H II}}
\mathcal{L}_{\rm BAO}$, where $\mathcal{L}_{\rm H II}=e^{-\chi^2_{\rm H II}/2}$,
to get the best fit set of parameters $\bf p_*$, and the 1, 2, and
3$\sigma$ contours defined as points where the likelihood 
equals $e^{-2.30/2}$, $e^{-6.17/2}$, and $e^{-11.8/2}$, respectively, 
of the maximum likelihood value. These contours are shown in Figs.\
4---6. Clearly, even adding currently available H II galaxy data
to the mix tightens the constraints. 

\section{Conclusion}
\label{conc}
We have used the starburst galaxy luminosity distance data of 
\cite{siegel2005} to constrain cosmological parameters in some dark
energy models. The resulting constraints are largely consistent with 
those that follow from other currently available data although they
are not as restrictive as the constraints derived from SNeIa, BAO,
and CMB anisotropy data.

The \cite{siegel2005} H II data are preliminary H II data. It is
gratifying that they result in interesting constraints on cosmological
parameters. We anticipate that new, near-future, H II galaxy data will
provide very useful constraints on cosmological parameters that will 
complement those derived using other data.

\acknowledgments
DM thanks Manolis Plionis, Glenn Horton-Smith, and Yun Chen for 
valuable suggestions. We acknowledge support from DOE grant DEFG030-99EP41093,
NSF grant AST-1109275, and from the SNSF (SCOPES grant No 128040).

\newpage
\begin{figure}[h]
\centering
\includegraphics[width=410pt,keepaspectratio=true]{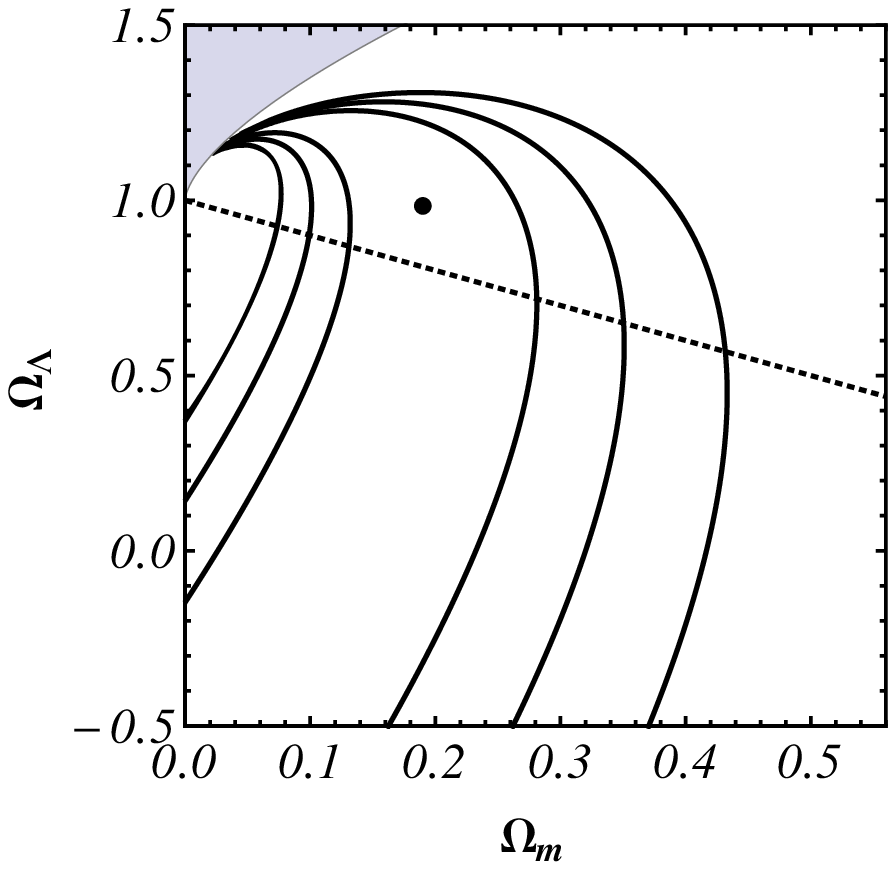}
% .: 0x0 pixel, 0dpi, 0.00x0.00 cm, bb=
 \caption{H II galaxy data 1, 2, and 3$\sigma$ confidence level contours 
in the ($\Omega_m, \Omega_\Lambda$) plane for the $\Lambda$CDM model. 
The dotted line corresponds to the spatially-flat $\Lambda$CDM case and
the shaded area in the upper left hand corner is the part of parameter space
without a big bang. The best-fit point with $\chi^2_{\rm min}=53.3$ is 
indicated by the solid black circle at $\Omega_m=0.19$ and 
$\Omega_\Lambda=0.98$.}
\label{fig: 1}
\end{figure}
\newpage
\begin{figure}[h]
 \centering
 \includegraphics[width=410pt,keepaspectratio=true]{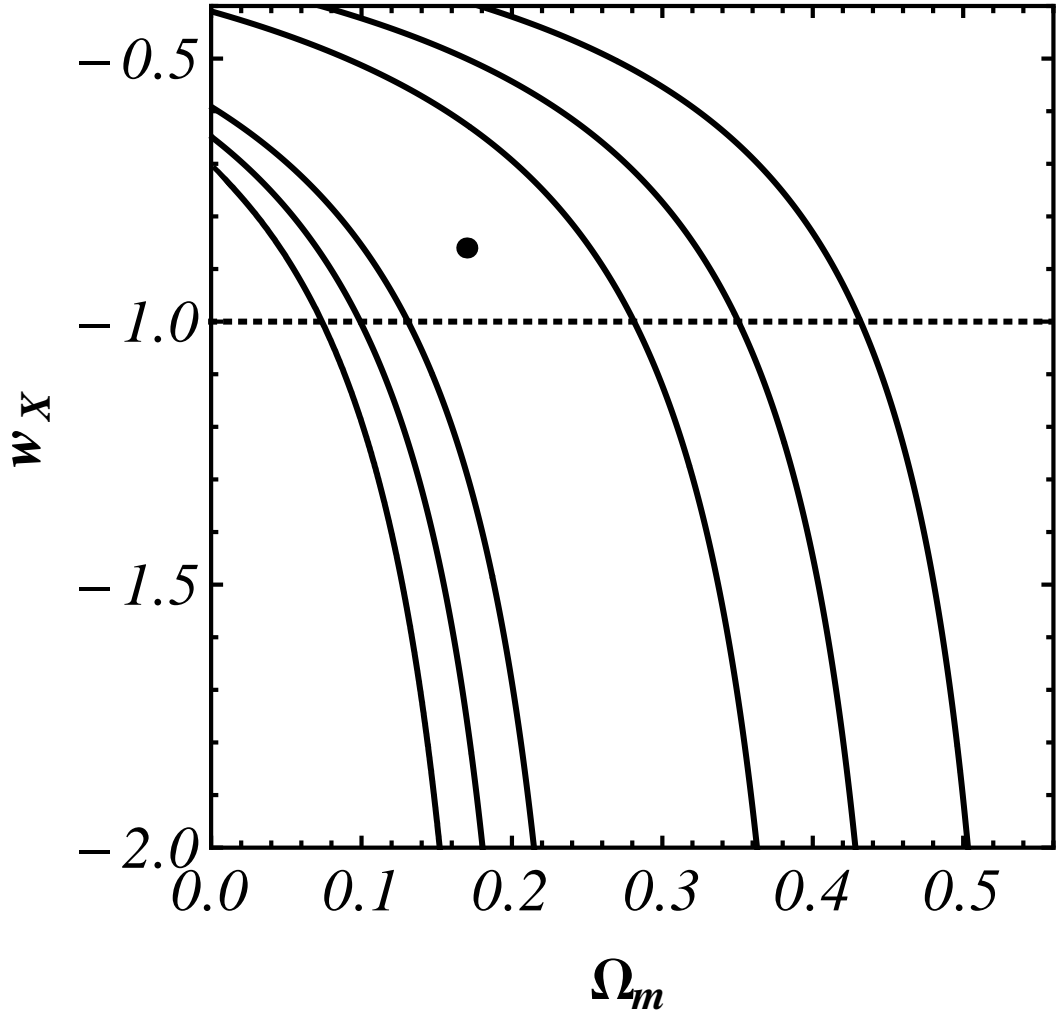}
% .: 0x0 pixel, 0dpi, 0.00x0.00 cm, bb=
 \caption{H II galaxy data 1, 2, and 3$\sigma$ confidence level contours 
in the ($\Omega_m, ~w_X$) plane for the spatially-flat XCDM parametrization. 
The dotted line corresponds to the spatially-flat $\Lambda$CDM case. 
The best-fit point with $\chi^2_{\rm min}=53.3$ is indicated by the 
solid black circle at $\Omega_m=0.17$ and $w_X=-0.86$.}
 \label{fig: 2}
\end{figure}
\newpage
\begin{figure}[h]
 \centering
 \includegraphics[width=410pt,keepaspectratio=true]{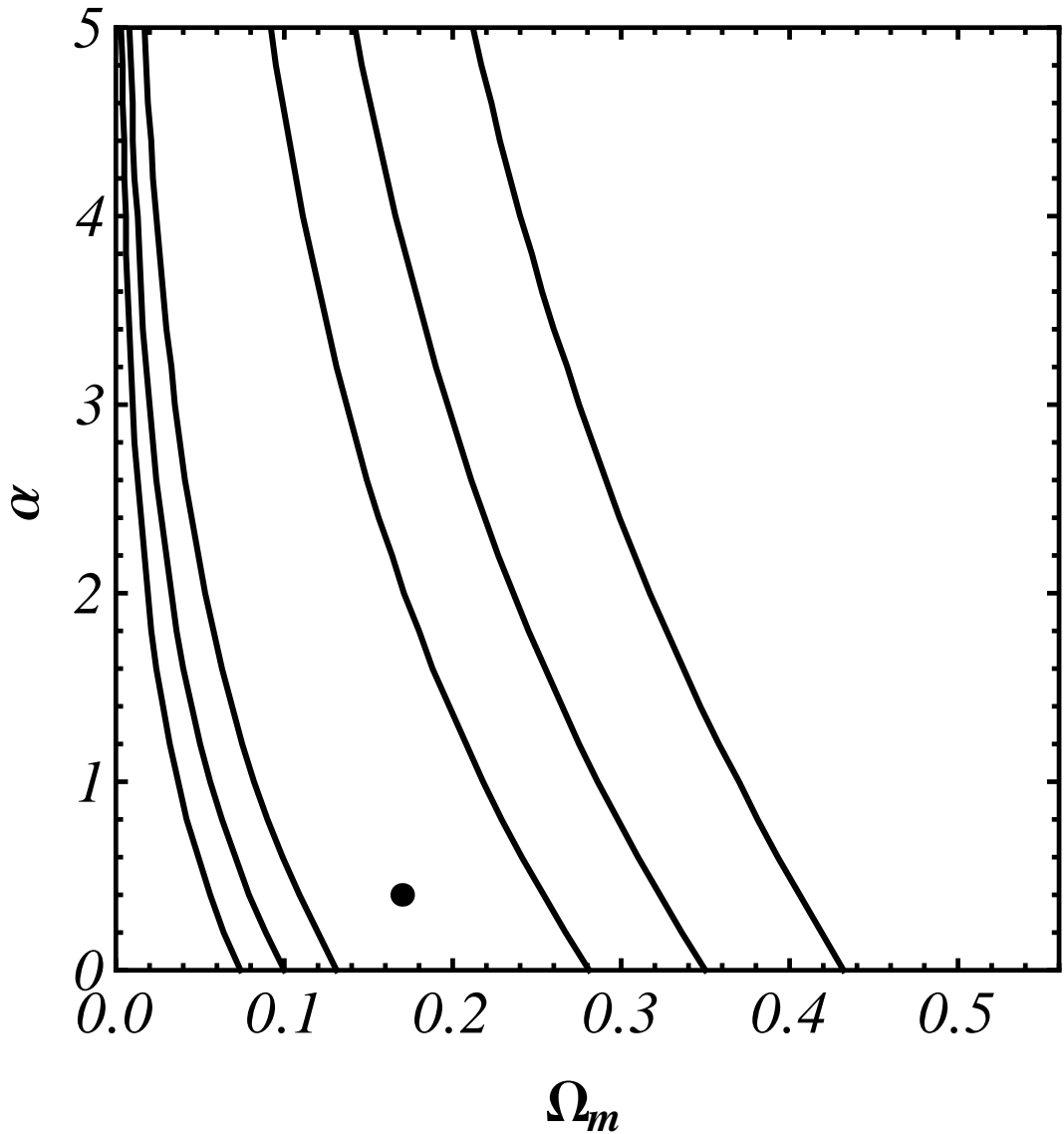}
% .: 0x0 pixel, 0dpi, 0.00x0.00 cm, bb=
 \caption{H II galaxy data 1, 2, and 3$\sigma$ confidence level contours 
in the ($\Omega_m, ~\alpha$) plane for the spatially-flat $\phi$CDM model. 
$\alpha=0$ corresponds to the spatially-flat $\Lambda$CDM case. 
The best-fit point with $\chi^2_{\rm min}=53.3$ is indicated by the solid 
black circle at $\Omega_m=0.17$ and $\alpha=0.39$.}
 \label{fig: 3}
\end{figure}
\newpage
\begin{figure}[h]
 \centering
 \includegraphics[width=410pt,keepaspectratio=true]{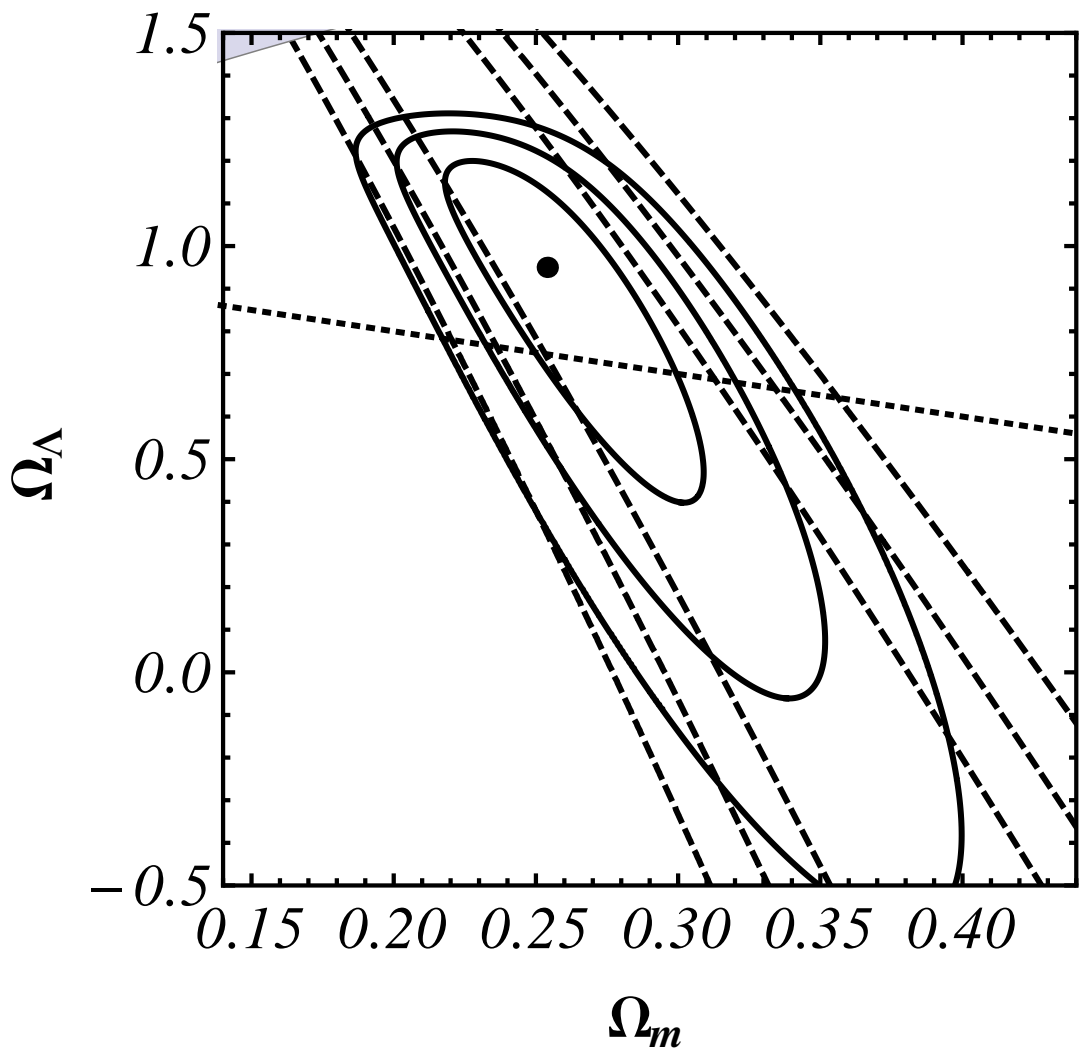}
% .: 0x0 pixel, 0dpi, 0.00x0.00 cm, bb=
 \caption{Joint H II galaxy and BAO data (solid lines) and BAO data only 
(dashed lines) 1, 2, and 3$\sigma$ confidence level contours in the 
($\Omega_m, \Omega_\Lambda$) plane for the $\Lambda$CDM model. 
Conventions and notation are as in Fig.\ 1. The best-fit point with 
$-2\log(\mathcal{L}_{\max})=55.2$ is indicated by the solid black circle at 
$\Omega_m=0.25$ and $\Omega_\Lambda=0.95$.}
 \label{fig: 4}
\end{figure}
\newpage
\begin{figure}[h]
 \centering
 \includegraphics[width=410pt,keepaspectratio=true]{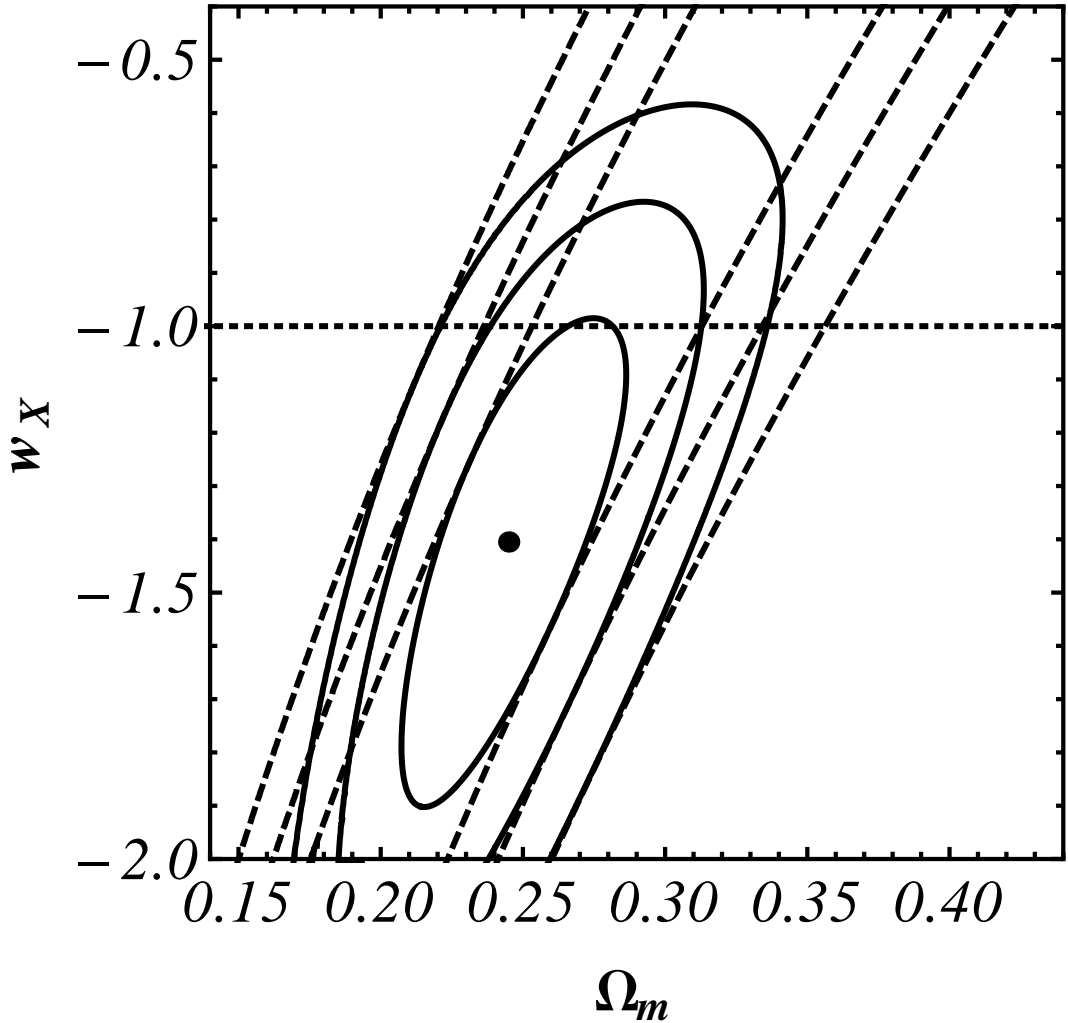}
% .: 0x0 pixel, 0dpi, 0.00x0.00 cm, bb=
 \caption{Joint H II galaxy and BAO data (solid lines) and BAO data only 
(dashed lines) 1, 2, and 3$\sigma$ confidence level contours in the 
($\Omega_m, ~w_X$) plane for the spatially-flat XCDM parametrization. 
Conventions and notation are as in Fig.\ 2. The best-fit point with
$-2\log\mathcal({L}_{\max})=53.5$ is indicated by the solid black circle 
at $\Omega_m=0.25$ and $w_x=-1.41$.
}
 \label{fig: 5}
\end{figure}
\newpage
\begin{figure}[b]
 \centering
 \includegraphics[width=410pt,keepaspectratio=true]{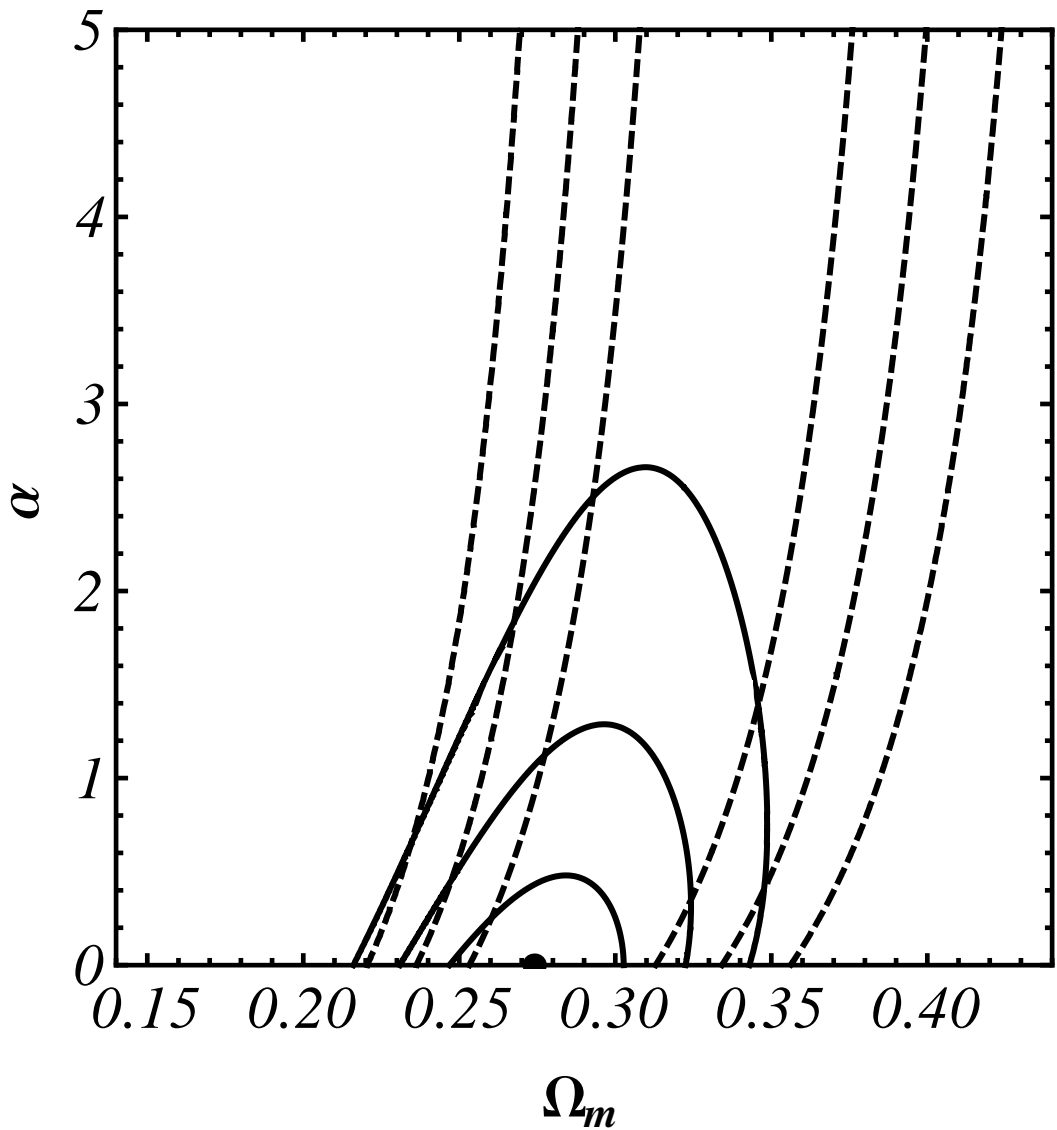}
% .: 0x0 pixel, 0dpi, 0.00x0.00 cm, bb=
 \caption{Joint H II galaxy and BAO data (solid lines) and BAO data only 
(dashed lines) 1, 2, and 3$\sigma$ confidence level contours in the 
($\Omega_m, ~\alpha$) plane for the spatially-flat $\phi$CDM model
Conventions and notation are as in Fig.\ 3. The best-fit point with
$-2\log(\mathcal{L}_{\max})=55.6$ is indicated by the solid black circle 
at $\Omega_m=0.27$ and $\alpha=0$.
}
 \label{fig: 6}
\end{figure}

\begin{table*}

\caption{\cite{siegel2005} H II starburst galaxy distance moduli and 
 uncertainties  }
\begin{center}
\begin{tabular}{l|c|c}
\label{tabl}
Galaxy & $z$ & $\mu_{\rm obs} \pm \sigma$ \\ 
\hline
\hline
Q0201-B13 & 2.17 & $47.49^{+2.10}_{-3.43}$ \\
Q1623-BX432 & 2.18 & $45.45^{+1.97}_{-3.07}$ \\
Q1623-MD107 & 2.54 & $44.82^{+0.31}_{-1.58}$ \\
Q1700-BX717 & 2.44 & $46.64^{+0.31}_{-1.58}$ \\
CDFa C1 & 3.11 & $45.77^{+0.31}_{-1.58}$ \\
Q0347-383 C5 & 3.23 & $47.12^{+0.44}_{-0.32}$ \\
B2 0902+343 C12 & 3.39 & $46.96^{+0.71}_{-0.81}$ \\
Q1422+231 D81 & 3.10 & $48.81^{+0.38}_{-0.40}$ \\
SSA22a-MD46 & 3.09 & $46.76^{+0.56}_{-0.51}$ \\
SSA22a-D3 & 3.07 & $49.71^{+0.43}_{-0.41}$ \\
DSF2237+116a C2 & 3.32 & $47.73^{+0.25}_{-0.25}$ \\
B2 0902+343 C6 & 3.09 & $45.22^{+1.38}_{-1.76}$ \\
MS1512-CB58 & 2.73 & $47.49^{+1.22}_{-1.57}$ \\
\end{tabular}
\end{center}
\end{table*}
\end{document}